\documentclass[apj]{emulateapj}
\usepackage{graphics,graphicx,times,natbib,longtable,placeins}

\def\mhalo{${\rm M_{halo}}$}
\def\mstar{${\rm M_{*}}$}
\def\msun{${\rm\,M_\odot}$}

\def\hst{{\it HST}}

\def\nar{New Astron. Rev.}

\def\spose#1{\hbox to 0pt{#1\hss}}
\def\lta{\mathrel{\spose{\lower 3pt\hbox{$\mathchar"218$}}
     \raise 2.0pt\hbox{$\mathchar"13C$}}}

\shorttitle{Dwarf Star Formation Histories}
\shortauthors{Guo et al.}

\begin{document}

\title{The Bursty Star Formation Histories of Low-mass Galaxies at 0.4$<$z$<$1
Revealed by Star Formation Rates Measured from H$\beta$ and FUV}

\author{Yicheng Guo$^{1}$, Marc Rafelski$^{2,3,13}$, S. M. Faber$^{1}$, David C.
Koo$^{1}$, Mark R. Krumholz$^{1,4}$, Jonathan R. Trump$^{5,6}$, S. P.
Willner$^{7}$, Ricardo Amor{\'i}n$^{8}$, Guillermo Barro$^{1,9}$, Eric F.
Bell$^{10}$, Jonathan P. Gardner$^{2}$, Eric Gawiser$^{11}$, Nimish P.
Hathi$^{12}$, Anton M. Koekemoer$^{13}$, Camilla Pacifici$^{2,3}$, Pablo G. P\'erez-Gonz\'alez$^{14}$,
Swara Ravindranath$^{13}$, Naveen Reddy$^{15}$, Harry I. Teplitz$^{16}$, Hassen
Yesuf$^{1}$}
\affil{$^1$ UCO/Lick Observatory, Department of Astronomy and Astrophysics, University of California, Santa Cruz, CA, USA; {\it ycguo@ucolick.org}}
\affil{$^2$ Goddard Space Flight Center, Code 665, Greenbelt, MD, USA}
\affil{$^3$ NASA Postdoctoral Program Fellow}
\affil{$^4$ Research School of Astronomy, and Astrophysics, Australian National University, Canberra, ACT 2611, Australia}
\affil{$^5$ Department of Astronomy and Astrophysics and Institute for Gravitation and the Cosmos, Pennsylvania State University, University Park, PA, USA}
\affil{$^6$ Hubble Fellow}
\affil{$^{7}$ Harvard-Smithsonian Center for Astrophysics, Cambridge, MA, USA}
\affil{$^8$ INAF-Osservatorio Astronomico di Roma, Monte Porzio Catone, Italy}
\affil{$^9$ Department of Astronomy, University of California, Berkeley, CA, USA}
\affil{$^{10}$ Department of Astronomy, University of Michigan, Ann Arbor, MI, USA}
\affil{$^{11}$ Department of Physics and Astronomy, Rutgers University, New Brunswick, NJ, USA }
\affil{$^{12}$ Aix Marseille Universit{\'e}, CNRS, LAM (Laboratoire d'Astrophysique de Marseille) UMR 7326, Marseille, France}
\affil{$^{13}$ Space Telescope Science Institute, Baltimore, MD, USA}
\affil{$^{14}$ Departamento de Astrof\'{\i}sica, Facultad de CC.  F\'{\i}sicas, Universidad Complutense de Madrid, E-28040 Madrid, Spain}
\affil{$^{15}$ Department of Physics and Astronomy, University of California, Riverside, CA, USA}
\affil{$^{16}$ Infrared Processing and Analysis Center, Caltech, Pasadena, CA 91125, USA}



\begin{abstract} 

We investigate the burstiness of star formation histories (SFHs) of galaxies at
$0.4<z<1$ by using the ratio of star formation rates (SFRs) measured from
H$\beta$ and FUV (1500 \AA) (H$\beta$--to--FUV ratio). Our sample contains 164
galaxies down to stellar mass (\mstar) of $10^{8.5}$\msun\ in the CANDELS
GOODS-N region, where Team Keck Redshift Survey Keck/DEIMOS spectroscopy and
\hst/WFC3 F275W images from CANDELS and Hubble Deep UV Legacy Survey are
available. When the {\it ratio} of H$\beta$- and FUV-derived SFRs is measured,
dust extinction correction is negligible (except for very dusty galaxies) with
the Calzetti attenuation curve. The H$\beta$--to--FUV ratio of our sample
increases with \mstar\ and SFR. The median ratio is $\sim$0.7 at
\mstar$\sim10^{8.5}$\msun\ (or SFR$\sim$0.5\msun/yr) and increases to $\sim$1 at
\mstar$\sim10^{10}$\msun\ (or SFR $\sim$10\msun/yr). At \mstar$<10^{9.5}$\msun,
our median H$\beta$--to--FUV ratio is lower than that of local galaxies at the
same \mstar, implying a redshift evolution.  Bursty SFH on a timescale of a few
tens of megayears on galactic scales provides a plausible explanation of our
results, and the importance of the burstiness increases as \mstar\ decreases.
Due to sample selection effects, our H$\beta$--to--FUV ratio may be an upper
limit of the true value of a complete sample, which strengthens our
conclusions. Other models, e.g., non-universal initial mass function or
stochastic star formation on star cluster scales, are unable to plausibly
explain our results. 

\end{abstract}

\section{Introduction}
\label{intro}

The star formation histories (SFHs) of galaxies with stellar masses (\mstar)
lower than $10^9$\msun\ (low-mass galaxies or dwarf galaxies) are expected to
be bursty. In such galaxies, supernova feedback following an intense star
formation (SF) episode can quickly heat and expel gas from them, resulting in a
temporary quenching of SF. New gas accretion and recycling of the expelled gas
then induce new SF. Therefore, the SFHs of dwarf galaxies in many models are
characterized by frequent bursts of SF and subsequent quick quenchings on a
time-scale of a few or tens of megayears
\citep[e.g.,][]{hopkins14fire,shens14,sparre15}.

Observationally, in their landmark paper, \citet{searle73} tentatively
concluded, by using broadband colors, that the bluest (and dwarf) galaxies
undergo ``intermittent and unusually intense bursts of SF.'' The ratio of star
formation rates (SFRs) measured from nebular emission lines and ultraviolet
(UV) continuum provides a more direct tool than broadband colors to explore
the burstiness of galaxy SFHs. Balmer line emission arises from the
recombination of gas ionized by O-stars, which have lifetimes of only a few
megayears. Thus, H$\alpha$ and H$\beta$ emissions trace SFR over a time scale
of a few megayears. On the other hand, UV photons come from both O- and
B-stars, which last for $\sim$100 Myr. Therefore, UV traces SFRs over that
timescale. As a result, a galaxy with an SFH that forms a significant fraction
of its stars in bursts separated by $\sim$5--100 Myr will spend short amounts
of time with very high H$\alpha$ and H$\beta$ luminosities followed by long
periods of low H$\alpha$ and H$\beta$ luminosities. The time variation of their
FUV luminosities is much smoother.  This effect does not alter the average
luminosity---in comparison to galaxies with steady star formation histories of
the same averaged value, a sample of bursty galaxies will still have the same
mean FUV, H$\alpha$, and H$\beta$ luminosity---but it can alter the
distribution, median, and mean values of the H$\alpha$ (or H$\beta$)--to--FUV
ratio \citep{ip04,fumagalli11,weisz12,slug1,slug2}. Thus, a measurement of this
distribution or its moments can be a sensitive diagnostic of the SFH with the
distribution of ratios depending on the intensity, duration, and separation of
the bursts.

In the local universe, a number of authors have reported H$\alpha$--to--FUV SFR
ratios lower than unity for low-mass (and hence low-SFR) or dwarf galaxies
\citep[e.g.,][]{sullivan00,bell01,boselli09,leejanice09,meurer09,weisz12}. The
effect becomes noticeable for star formation rates below $\sim$0.1
M$_\odot$~yr$^{-1}$ and is absent in galaxies with higher star formation rates.
For a sample of galaxies, an average H$\alpha$--to--FUV ratio lower than unity
indicates that these galaxies are preferentially observed during their
subsequent quenching following SF bursts. 

Besides the bursty SFHs on galactic-wide scales, two other possibilities are
often used to explain the observed lower-than-unity H$\alpha$--to--FUV ratios:
non-universal initial mass function (IMF) and stochastic SF on small (star or
star cluster) scales. Non-universal IMF theories assume that the IMFs of
galaxies depend on the properties of galaxies. One particular scenario, the
integrated galactic initial mass function (IGIMF), predicts that the actual IMF
is steeper than the canonical IMF and steepens with the decrease of the total
SFR of galaxies \citep{weidner05,pa07,pa09,weidner11}. This scenario assumes
that (1) all stars form in clusters \citep[e.g.,][]{pa07}; (2) the cluster mass
function is a power law; (3) the mass of the most massive star cluster (i.e.,
the truncation mass of the power law) is a function of the total SFR of a
galaxy; and (4) the mass of the most massive star in a star cluster is a
function of the mass of the cluster. As a result, the chance of forming a
massive star in low-SFR galaxies is lower than in high-SFR galaxies.
Consequently, low-SFR galaxies have steeper IMFs than high-SFR galaxies. Due to
the lack of massive stars to ionize hydrogen, low-SFR galaxies, therefore, have
lower H$\alpha$--to--FUV ratio. \citet{leejanice09} found that the IGIMF model
is able to account for the observed H$\alpha$--to--FUV ratio in their Local
Volume galaxy sample, but they also pointed out that a critical test of the
IGIMF model would be the scatter of the H$\alpha$--to--FUV ratio.
\citet{fumagalli11} and \citet[][W12]{weisz12} compared the IGIMF model to
their samples and found that the scatter predicted by IGIMF is not compatible
with observations. In addition, observations of individual low-mass star
clusters have failed to detect the deficiency in H$\alpha$ predicted by IGIMF
models, and appear fully consistent with a universal, randomly sampled IMF
\citep{andrews13,andrews14}. While \citet{andrews13,andrews14} were testing the
mass of the most massive star in a star cluster (i.e., Assumption (4) listed
above), \citet{pa13} tested Assumption (3) --- the mass of the most massive
star cluster --- and found that the masses of the most massive young star
clusters in M33 decrease with increasing galactocentric radius, supporting a
non-universal truncation mass of the cluster mass function. \citet{weidner14}
showed that some weaker versions of the IGIMF hypothesis
\citep[e.g.,][]{weidner06} remain consistent with the Andrews et al. data, but
the fact that the observations are also consistent with a universal IMF implies
that they also provide no support for the hypothesis of non-universality. 


Stochastic SF on small scales
\citep[e.g.,][]{cervino03,cervino04,fumagalli11,eldridge12,slug1,cervino13} can
also produce low H$\alpha$--to--FUV ratio for low-SFR galaxies. Sometimes,
bursty SFH is also considered as a stochastic process, but in this paper we use
the following distinction: burstiness is caused by phenomena on galaxy scales
(e.g., feedback, gas accretion, and merger, etc.), while stochasticity occurs
on star or star cluster scale. The stochasticity considered in our paper is the
stochastic sampling of IMFs and SFHs. Even for a universal IMF, when the SFR is
low, the IMF would not be fully sampled. Instead, the random sampling would
bias against very massive stars because of their low formation probability,
resulting in an actually steeper IMF. 

Similarly, an SFH would not be fully sampled when the SFR is low, because a
time-averaged continuous (the undersampled) SFR would actually consist of
several small ``bursts'' associated with the formation of new star clusters.
Mathematically , the SFH is not fully sampled over a time interval $T$ if
$\frac{1}{\langle M_c\rangle} \int_0^T \mbox{SFR}(t) dt \lesssim 1$, where
$\langle M_c\rangle$ is the expected mass of a single star cluster. That is,
the SFH is not fully sampled over a specified time interval if the expected
number of star clusters formed over that time interval is of order 1 or
fewer.\footnote{We pause to point out a subtlety in terminology: as used in
this statement, a star cluster is simply defined as a collection of stars that
formed at a single instant in time, without regard to whether it is
gravitationally bound or relaxed, and the implicit assumption we make is that
the great majority of stars form in such temporally coherent structures.  While
most stars do not form in gravitationally bound clusters
\citep[e.g.,][]{lada03,johnson16}, observations indicate that most stars do
form in clusters by the weaker definition of a cluster that is relevant for our
purposes \citep[e.g.,][]{lada03,fall12}. For more discussion on this point, see
\citet{krumholz14}.}
\citet{slug1} developed a simulation tool SLUG (Stochastically Lighting Up
Galaxies) to study the stochasticity in SF and its effects on SFR indicators
and stellar population \citep{slug2,slug3}. Using SLUG, \citet{fumagalli11}
showed that stochasticity is able to explain the observed low
H$\alpha$--to--FUV ratio in local dwarf galaxies
\citep{boselli09,leejanice09,meurer09}.

Although evidence of bursty SFHs of dwarf galaxies in the nearby universe has
been present through the H$\alpha$--to--FUV ratio in some of the above studies
\citep[e.g.,][W12]{leejanice09}, similar observations have not been conducted
beyond the local universe because of the lack of both deep UV and deep
optical/IR spectroscopy data. Low-mass galaxies at higher redshifts are
expected to have similar or even burstier SFHs \citep[e.g.,][]{dominguez15}
compared to their local counterparts. Recently, \citet{kurczynski16} found no
statistically significant increase of the intrinsic scatter in the SFR--\mstar\
relation at low masses at $0.5<z<3.0$, which seems to indicate a gradual
assembly of galaxy masses. Their SFRs, however, were measured from SED-fitting
of broadband photometry, which traces SFR on timescales of 100 Myr. It,
therefore, remains possible that the SF burstiness (indicated by the intrinsic
scatter) increases at low masses on timescales shorter than $\sim$100 Myr. On
the other hand, both non-universal IMF and stochastic SF are phenomena on the
scales of star clusters and, therefore, independent of redshift. Studying the
redshift evolution of the H$\alpha$--to--FUV ratio would shed a light on the
importance of the burstiness of SFHs of low-mass galaxies. 

In this paper, to investigate the burstiness of the SFHs beyond the local
universe, we use optical spectroscopy from the Team Keck Redshift Survey
\citep[TKRS,][]{wirth04} and UV imaging from CANDELS
\citep{candelsoverview,candelshst} and the Hubble Deep UV (HDUV) Legacy Survey
(HST-GO-13871, PI. Oesch) to measure the ratio of H$\beta$ and FUV (1500 \AA)
derived SFRs of galaxies at $0.4<z<1$ in the CANDELS GOODS-N field. 
The advantage of using H$\beta$ instead of H$\alpha$ is that the dust
extinction effects on H$\beta$ and FUV almost cancel each other out, because
they have very similar extinction (assuming the Calzetti law and an extra
nebular extinction). Therefore, generally, no extinction correction is needed
when measuring the intrinsic ratio of SFR$_{H\beta}$ and SFR$_{1500\AA}$ except
for very dusty galaxies (see Section \ref{ratio} and \ref{discussion:dust}).

We adopt a flat ${\rm \Lambda CDM}$ cosmology with $\Omega_m=0.3$,
$\Omega_{\Lambda}=0.7$, and the Hubble constant $h\equiv H_0/100\ {\rm
km~s^{-1}~Mpc^{-1}}=0.70$. We use the AB magnitude scale \citep{oke74} and a
\citet{chabrier03} IMF.

\begin{figure}[htbp]
\center{\hspace*{-0.75cm}\includegraphics[scale=0.31, angle=0]{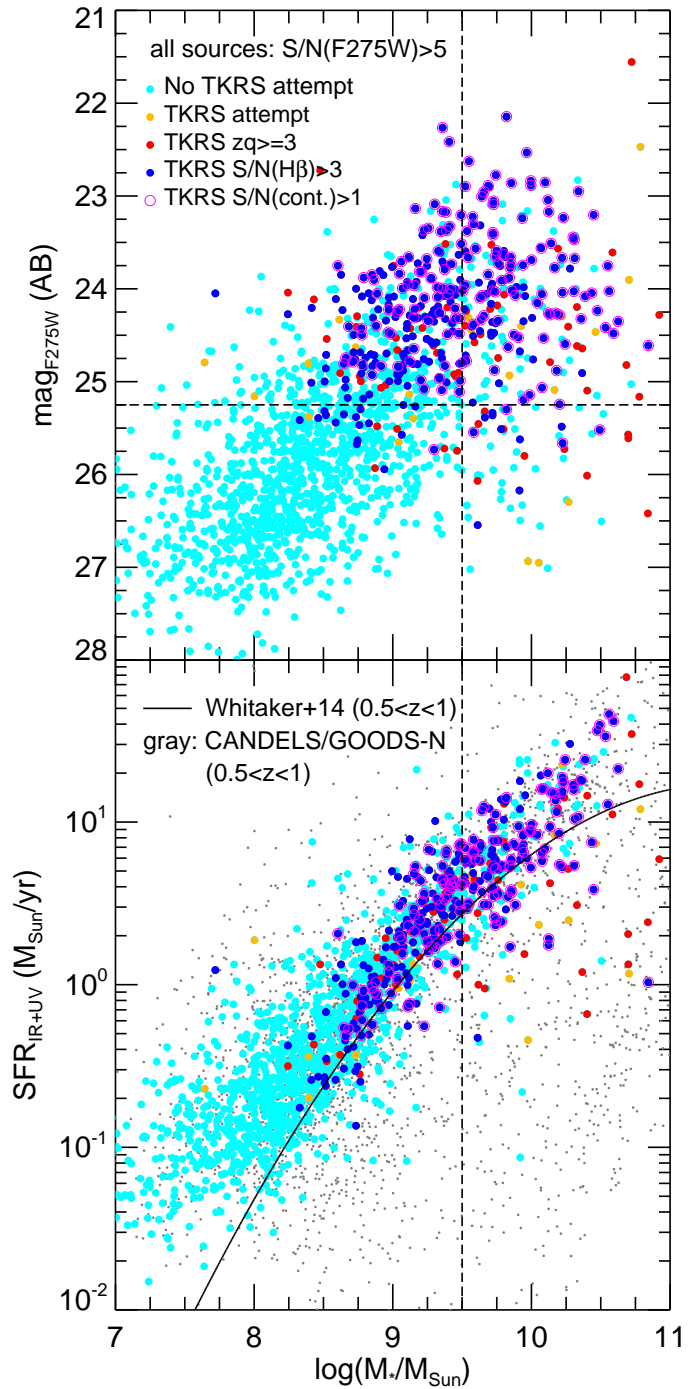}}

\caption[]{Sample selection. {\it Upper}: F275W magnitude--\mstar\ relation for
GOODS-N galaxies with F275W S/N$>$5. Galaxies with no TKRS attempt are cyan.
Those with TKRS attempt, but no secure redshift obtained, are yellow.  We only
measure H$\beta$ fluxes for galaxies with secure TKRS redshifts (red), and only
select those with H$\beta$ S/N$>$3 (blue) {\it and} continuum S/N$>$1/pixel
(blue+purple circle) into our sample. {\it Lower}: SFR--\mstar\ relation.
Galaxies are color coded as the upper panel. All CANDELS GOODS-N galaxies at
$0.5<z<1$ are overplotted as gray points. The best-fit relation of
\citet{whitaker14} is also plotted as a reference (solid line). The method of
measuring SFR$_{tot}$ is described in Section \ref{data}.

\label{fig:sample}}
\end{figure}

\section{Data and Sample}
\label{data}

TKRS observed a magnitude-limited sample of 2911 objects to $R<24.4$ AB and
yielded secure redshifts for 1440 galaxies and active galactic nuclei.  TKRS
used the 600 lines ${\rm mm^{-1}}$ grating blazed at 7500\AA. The central
wavelength was set at 7200\AA, providing a nominal spectral coverage of
4600--9800\AA\ at a FWHM resolution of $\sim$3.5\AA\ and $R\sim2500$. Each slit
mask was observed for a total on-source integration time of 3600 seconds.

We follow \citet{trump13} and \citet{ycguo16mzr} to measure the flux of
H$\beta$. First, a continuum is fitted across the emission-line region by
splining the 50-pixel smoothed continuum. Then, a Gaussian function is fitted
to the continuum-subtracted flux in the wavelength region of H$\beta$. The
emission-line intensity is computed as the area under the best-fit Gaussian in
the line wavelength region. The flux calibration is done by scaling the
continuum flux density to match the best-fit stellar population model of the
broadband spectral energy distribution (SED) from the CANDELS GOODS-N
multiwavelength catalog (Barro et al., in preparation, see
\citet{ycguo13goodss} and \citet{galametz13uds} for details).  The advantage of
this calibration is that using the broadband flux of whole galaxies corrects
for the slit-loss effect, under a (reasonable) assumption that the continuum
(stellar flux) and emission line have the same spatial distribution. To obtain
a reliable flux calibration, we only use galaxies with continuum
signal-to-noise ratio (S/N)$>$1/pixel. We also require H$\beta$ S/N$>$3 to use
reliable H$\beta$ flux measurements.

The raw H$\beta$ equivalent widths are corrected for stellar absorption
according to galaxy ages measured through SED-fitting. As part of the CANDELS
SED-fitting effort, each galaxy in our sample has been fitted by 12 SED-fitting
codes, which have different combinations of synthetic stellar population
models, SFHs, fitting methods, etc. (see \citet{santini15} for details). For
each galaxy, we use the median of the 12 best-fit ages as its age to measure
the stellar absorption from the models of \citet{bc03}. The average
light-weighted age of galaxies in our sample is almost constant ($\sim$1 Gyr)
from $10^{8.5}$ to $10^{10.5}$\msun, corresponding to an average stellar
absorption equivalent width of $\sim$2.5 \AA. We also try to apply a fixed
equivalent width of 1 \AA\ \citep{cowie08,zahid11,henry13a} to all galaxies.
The smaller stellar absorption equivalent width does not change our results and
conclusions, because our galaxies have very large H$\beta$ equivalent widths (a
median of $\sim$15 \AA).

The UV luminosities of galaxies in our sample are measured from \hst/WFC3 F275W
images of CANDELS UV and HDUV. CANDELS UV covers the CANDELS GOODS-N Deep
region in F275W at an approximately four-orbit depth. HDUV intends to add four
orbits of observations to each of eight pointings in F275W in the GOODS-N Deep
region. We use the first public release (v0.5) of
HDUV\footnote{http://www.astro.yale.edu/hduv/data.html}, which includes five
out of the eight new deeper pointings from HDUV, plus all of the CANDELS UV
exposures. 

The UV photometry is measured in a similar fashion as done for the Ultraviolet
Hubble Ultra Deep Field (UVUDF) \citep{rafelski15}. SExtractor is run on F275W
and ACS/F435W images in dual-image mode with the F435W image as the detection
image. An aperture correction is then determined from the F435W band to match
the small F435W-band apertures to the larger F160W-band apertures used in the
rest of the CANDELS catalog of Barro et al. (in preparation). We only include
galaxies with F275W S/N$>$5 in our analyses. There are about 1700 galaxies with
F275W S/N$>$5 at $0.4 < z < 1.0$ in our UV catalog. Our final sample with both
H$\beta$ S/N$>$3 and F275W S/N$>$5 contains 164 galaxies after excluding
X-ray-detected sources.

We use FAST \citep{kriek09fast} to fit the \citet{bc03} models to the CANDELS
GOODS-N multiwavelength catalog to measure \mstar. The total SFRs (SFR$_{tot}$)
are measured by combining dust-uncorrected rest-frame UV (2800 \AA) and IR
luminosities, if the latter (measured from {\it Spitzer}/MIPS and/or {\it
Herschel}) is available, by following \citet{kennicutt98}. For these galaxies,
SFR$_{tot}$ = SFR$_{UV+IR}$. If no IR measurement is available, SFR$_{tot}$ is
measured through rest-frame UV (2800 \AA) corrected for dust extinction
measured from SED-fitting, i.e., SFR$_{tot}$ = SFR$_{NUV,corr}$. We refer
readers to \citet{barro11b, barro13a} for the details. Because F275W photometry
is not used in the SFR$_{tot}$ measurement, SFR$_{tot}$ is independent of our
H$\beta$--to--FUV ratios.

Figure \ref{fig:sample} shows that at \mstar$>10^{9.5}$\msun, our sample
(galaxies with H$\beta$ S/N$>$3 {\it and} continuum S/N$>$1) is representative
of galaxies at $0.4<z<1$ in terms of F275W magnitude and total SFR. At
$10^{8.5}<$\mstar$<10^{9.5}$\msun, our sample is biased toward UV-bright
galaxies ({\it upper} panel of Figure \ref{fig:sample}). This bias is
introduced by the TKRS magnitude limit ($R<24.4\ AB$) and our H$\beta$ S/N
requirement, the latter of which also biases our sample toward H$\beta$-bright
galaxies. For our goal of measuring the H$\beta$--to--FUV ratio,
we will demonstrate later that the actual effect of our sample selection
(combining both H$\beta$ S/N$>$3 and F275W S/N$>$5) would bias our results
against low H$\beta$--to--FUV for low-mass galaxies (Section
\ref{discussion:bias}).

\begin{figure}[htbp] 
\center{\hspace*{-0.5cm}\includegraphics[scale=0.29, angle=0]{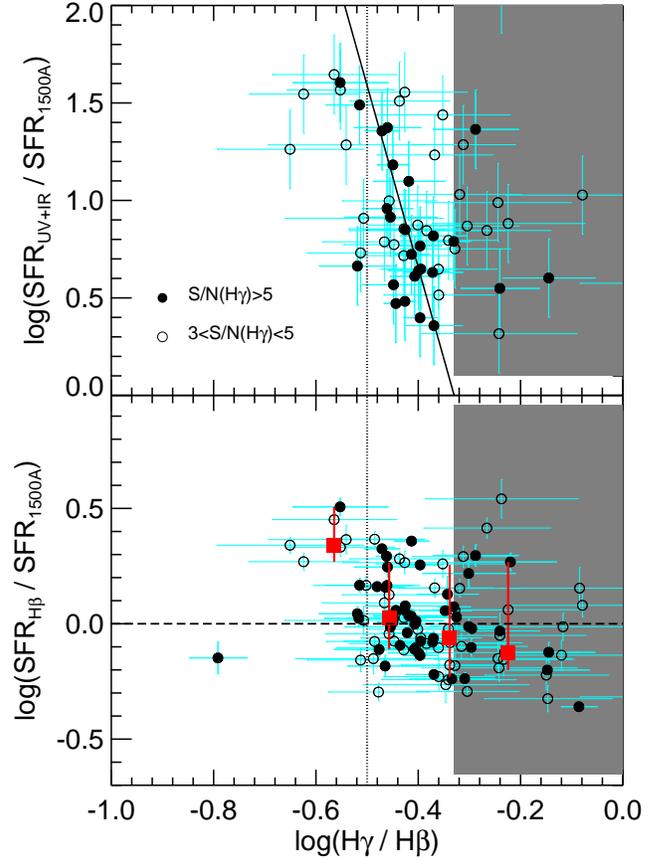}}

\caption[]{{\it Upper}: total--to--FUV SFR ratio as a function of the Balmer
decrements (${\rm H\gamma/H\beta}$) for galaxies with S/N(H$\gamma$)$>$3 (open)
and 5 (filled). All galaxies here have IR detections. The solid line shows a
fit to galaxies with S/N(H$\gamma$)$>$5. The shaded area is where ${\rm
H\gamma/H\beta}$$>$0.468, the intrinsic Balmer flux ratio. {\it Lower}:
H$\beta$--to--FUV ratio (uncorrected for dust effect) as a function of the
Balmer decrements. The median (red squares) and 16th and 84th percentiles
(error bars) are measured for both S/N(H$\gamma$)$>$3 (open) and $>$5(filled)
galaxies. The lower panel has more points than the upper panel, because not
every galaxy in the lower panel has an IR measurement.

\label{fig:dust}}
\end{figure}

\section{Measuring the H$\beta$--to--FUV Ratio}
\label{ratio}

${\rm SFR_{H\beta}}$ and ${\rm SFR_{1500\AA}}$ are
calculated\footnote{Throughout the paper, we assume that all the 1500 \AA\
emission is purely stellar, which is a common assumption in most studies. In
very young systems with age $\lesssim$1 Myr, however, nebular emission actually
contributes $\sim$20\% to the 1500 \AA\ emission \citep{reines10}. However,
this contribution quickly drops to $\leq$5\% as the systems become 5 Myr old.
We, therefore, conclude that the nebular contribution to 1500 \AA\ is
negligible in our sample.} through the formula of
\citet{kennicutt12}\footnote{Using the recipes of \citet{kennicutt98} results
in a systematically lower H$\beta$--to--FUV ratio by 0.03 dex.}
with an intrinsic H$\alpha$/H$\beta$ ratio of 2.86 (added as the last term in
the right-hand side of Equation (1) below): 

\begin{eqnarray} 
\label{eq:sfrhb} 
{\rm log(SFR_{H\beta}) = log(L_{H\beta}) - 41.27 + log(2.86)}  \\
{\rm log(SFR_{1500\AA}) = log(L_{1500\AA}) - 43.35},
\end{eqnarray} 
\noindent where both ${\rm L_{H\beta}}$ and ${\rm L_{1500\AA}}$ ($\nu L_{\nu}$) are in
units of ${\rm erg\ s^{-1}}$. 
Both formulae have been converted into a Chabrier IMF and assume a solar
metallicity. 

Low-mass galaxies tend to have subsolar metallicity. The average 
gas-phase metallicity of $10^{8.5}$\msun\ galaxies at $z\sim0.7$ is about
0.5$Z_\odot$ \citep{ycguo16mzr}. \citet{castellano14} showed that using
rest-frame UV tends to underestimate the SFRs of subsolar metallicity 
(gas-phase $\sim$0.3$Z_\odot$) Lyman Break Galaxies at $z\sim3$ by a factor of
a few compared to using SED-fitting or using nebular emission lines. If their
results also hold for low-redshift low-mass galaxies, using a UV SFR tracer
assuming solar metallicity would artificially increase the H$\beta$--to--FUV
ratio for low-mass galaxies, implying that the intrinsic burstiness of low-mass
galaxies is actually stronger than our results.

\begin{figure*}[htbp] 
\includegraphics[scale=0.3, angle=0, clip]{./fig3a_proof.ps}
\includegraphics[scale=0.3, angle=0, clip]{./fig3b_proof.ps}

\caption[]{{\it Left}: H$\beta$--to--FUV ratio as a function of \mstar. Black
circles with gray error bars are the best-measured ratios and their
uncertainties. For most of our galaxies, the best-measured ratios are not
corrected for dust extinction (see Section \ref{ratio}). However, for very
dusty galaxies (${\rm SFR_{UV+IR}/SFR_{FUV}}$$>$20.0), the best ratios are the
ones corrected for dust extinction. As a reference, for such galaxies, their
uncorrected ratios are shown as purple stars with purple straight lines
connected to their corrected (best) values. Red squares with error bars show
the median and 16th and 84th percentiles of each \mstar\ bin. 
Black solid and dotted lines show the best-fit linear relation and its 95\%
confidence level. The data of W12 at $z=0$ are shown as the blue triangles, and
their mean is shown by the blue line. {\it Right}: Similar to the left, but
showing H$\beta$--to--FUV ratio as a function of total SFR (${\rm SFR_{tot}}$).
Results of W12 at $z=0$ are also shown (blue triangles and blue line).  Two
galaxies with H$\beta$--to--FUV larger than 0.5 dex are not shown in the
figure, but are still included when calculating the median, percentiles, and
the best-fit relation.

\label{fig:result}}
\end{figure*}

To calculate the intrinsic SFRs, both ${\rm L_{H\beta}}$ and ${\rm
L_{1500\AA}}$ should be corrected for dust extinction. A shortcut, however,
exists when calculating the {\it ratio} of ${\rm SFR_{H\beta}}$ and ${\rm
SFR_{1500\AA}}$: extinction correction is negligible, except for very dusty
galaxies, because the extinction of H$\beta$ and FUV almost cancel each other
out by coincidence. This is an advantage of using H$\beta$ instead of H$\alpha$
in our study. 

To illustrate the shortcut, we express the intrinsic H$\beta$--to--FUV ratio as
the following: \small \begin{eqnarray} \label{eq:sfrratio5} \left(\frac{\rm
SFR_{H\beta}}{\rm SFR_{1500\AA}}\right)_{int} & = & \left(\frac{\rm
SFR_{H\beta}}{\rm SFR_{1500\AA}}\right)_{obs} \times
\left(\frac{10^{0.4A(H\beta)}}{10^{0.4A(FUV)}}\right) \\ & = & \left(\frac{\rm
SFR_{H\beta}}{\rm SFR_{1500\AA}}\right)_{obs} \times
10^{0.4(A(H\beta)-A(FUV))}, \nonumber \end{eqnarray} \normalsize \noindent
where $A(H\beta)$ and $A(FUV)$ are the attenuation of H$\beta$ and FUV. Their
difference is: \small \begin{eqnarray} \label{eq:av} A(H\beta) - A(FUV) =
E(B-V)\left[\frac{k(4861\AA)}{0.44} - k(1500\AA)\right], \end{eqnarray}
\normalsize where the factor of 0.44 is used, because
$E(B-V)_{stellar}=0.44\times E(B-V)_{gas}$ \citep{calzetti00}. For the Calzetti
attenuation curve \citep{calzetti00}, \small \begin{eqnarray} \label{eq:dust}
k(4861\AA)/0.44 - k(1500\AA) \sim 0.1.  \end{eqnarray} \normalsize \noindent
Therefore, the factor in Equation (3) to correct for dust extinction is
$10^{0.4(A(H\beta)-A(FUV))} \sim 10^{0.04E(B-V)_{stellar}}$. For galaxies with
$E(B-V)_{stellar} < 1$, the dust extinction would bias the observed
H$\beta$--to--FUV ratio from the intrinsic one toward the low-value side only
by $<$0.04 dex. Since, on average, the dust extinction of galaxies with
$10^{8.5}$\msun$<$\mstar$<10^{10.0}$\msun\ is smaller than $E(B-V)_{stellar} =
0.2$ \citep{dominguez13}, this systematic error is negligible compared to other
sources of uncertainty.

This shortcut is of course only valid when galaxies have (1) the Calzetti
attenuation curve and (2) $E(B-V)_{stellar}=0.44\times E(B-V)_{gas}$ (or
$E(B-V)_{gas}=2.27\times E(B-V)_{stellar}$). There are other attenuation curves
in the literature, e.g., MW\footnote{The shortcut is also valid for the MW
attenuation curve, but with a slightly higher systematic offset. In this case,
the dust extinction would bias the observed H$\beta$--to--FUV ratio from the
intrinsic one toward the low-value side by only $<0.07$ dex for galaxies with
$E(B-V)_{stellar} < 1$. It is also important to note that the relation
$E(B-V)_{gas}=2.27\times E(B-V)_{stellar}$ in \citet{calzetti00} is derived
based on the assumption that a Galactic extinction curve applies to the nebular
emission.}, LMC, SMC, and recently MOSDEF \citep{reddy15,reddy16}.  Moreover,
several studies \citep[e.g.,][]{reddy10,wild11,wuyts13,reddy15} suggest that
the extra attenuation of nebular emission, compared to the stellar attenuation,
is smaller than what we use here ($E(B-V)_{gas}=2.27\times E(B-V)_{stellar}$).
The extra attenuation depends on the geometry \citep{wild11}, SFR
\citep{reddy15}, and/or specific star formation rate (sSFR)\citep{debarros15}
of the galaxies. For an attenuation curve other than the Calzetti curve or a
smaller-than-2.27 factor for the extra nebular attenuation, the extinction
correction factor $10^{0.4(A(H\beta)-A(FUV))}$ in Equation (3) would be
significantly smaller than unity, resulting in a non-negligible extinction
effect to decrease the observed (dust uncorrected) H$\beta$--to--FUV ratio (see
Section \ref{discussion:dust}).

The negligible effect of dust extinction correction in our sample is
demonstrated in Figure \ref{fig:dust}. For a small portion of our sample, we
measure the Balmer decrements through the flux ratio of H$\gamma$/H$\beta$. For
galaxies with H$\gamma$/H$\beta$$\geq$0.32 (corresponding to $E(B-V)\leq0.8$,
the vertical dotted line in Figure \ref{fig:dust}), the dust-uncorrected
H$\beta$--to--FUV ratio has no dependence on H$\gamma$/H$\beta$ \citep[see][for
similar results]{zeimann14}. Only for very dusty galaxies
(H$\gamma$/H$\beta$$\leq$0.32, i.e., galaxies with ${\rm log(H\gamma/H\beta)
\lesssim -0.5}$ in the lower panel of Figure \ref{fig:dust}), is the
dust-uncorrected H$\beta$--to--FUV ratio significantly larger than unity,
indicating that the dust affects FUV significantly more than it affects
H$\beta$. 
For these galaxies, an extinction correction is needed to measure the
H$\beta$--to--FUV ratio.

The ration of ${\rm SFR_{UV+IR}/SFR_{1500\AA}}$ can serve as a good dust
extinction indicator when H$\gamma$/H$\beta$ measurement is lacking. As shown
by the upper panel of Figure \ref{fig:dust}, the Balmer decrements show a
correlation with ${\rm SFR_{UV+IR}/SFR_{1500\AA}}$.

We only make a dust correction for galaxies with ${\rm
SFR_{UV+IR}/SFR_{1500\AA}} > 20$ (corresponding to
H$\gamma$/H$\beta$$\lesssim$0.32).  We use the best-fit relation in the upper
panel of Figure \ref{fig:dust} to infer their H$\gamma$/H$\beta$ ratios and
correct for the extinction for ${\rm SFR_{H\beta}}$. We then use ${\rm
SFR_{UV+IR}}$ as the dust-corrected UV SFR. For these galaxies the
dust-corrected H$\beta$--to--FUV ratio is, therefore, the ratio of ${\rm
SFR_{UV+IR}}$ and dust-corrected ${\rm SFR_{H\beta}}$. In our analyses, we use
the dust-corrected ratios as the best-measured ratios for these very dusty
galaxies, while for other galaxies, whose ${\rm SFR_{UV+IR}/SFR_{1500\AA}}<20$,
we use the uncorrected H$\beta$--to--FUV ratios as their best ratios. Only
$\sim$10\% of our entire sample are very dusty and, therefore, need the
dust-corrected ratios, and they all have \mstar$>10^{9.5}$\msun\ (purple
asterisks in the left panel of Figure \ref{fig:result}).

Some galaxies in our sample have H$\gamma$/H$\beta$$\geq$0.468 (the shaded area
in Figure \ref{fig:dust}), which seems to imply that not all galaxies in our
sample follow the assumptions of Case B recombination, a requirement of the
Kennicutt SFR recipes used in our analyses. We believe that these galaxies are
scattered into the shaded area by the large line-ratio uncertainties.
Considering the uncertainties (horizontal error bars), 12\% of the galaxies
with H$\gamma$ measurement in our sample have H$\gamma$/H$\beta$ larger than
0.468 by more than 1 $\sigma$, and only 4\% are larger than 2$\sigma$. This
distribution is broadly consistent with a Gaussian distribution with the mean
of 0.468, where 16\% (and 2.5\%) of the galaxies should be deviated from the
mean (0.468) by more than 1$\sigma$ (and 2$\sigma$). We, therefore, argue that
the assumption of Case B recombination is not statistically invalid for our
sample.

\begin{figure*}[htbp]
\includegraphics[scale=0.3, angle=0, clip]{./fig4a.ps}
\includegraphics[scale=0.3, angle=0, clip]{./fig4b.ps}

\caption[]{H$\beta$--to--FUV ratio as a function of the observed F275W
magnitude (left) and H$\beta$ flux (right). In each panel, our sample is
divided into two \mstar\ bins: \mstar$>10^{9.5}$\msun\ (red) and
\mstar$<10^{9.5}$\msun\ (blue). A solid line with the same color is the best
linear fit to the galaxies in each \mstar\ bin, and the dotted lines with the
same color show the 95\% confidence level of the fit.

\label{fig:bias}}
\end{figure*}

\section{Results}
\label{results}

Figure \ref{fig:result} shows that the H$\beta$--to--FUV ratios of the galaxies
in our sample increase with their \mstar. The median ratio is about 0.7 for
galaxies with \mstar$\sim10^{8.5}$\msun\ and increases to 1.0 for galaxies with
\mstar$>10^{9.5}$\msun. The trend is broadly consistent with the expectation
that the SFHs of lower-mass galaxies are burstier than those of massive
galaxies. The figure also shows that the H$\beta$--to--FUV ratios increase with
SFR$_{tot}$. The median ratio is about 0.7 for galaxies with SFR$<$1\msun/yr
and increases to $\sim$1 for galaxies with SFR$>$10\msun/yr. 

Statistical tests show that the correlation between H$\beta$--to--FUV ratio and
\mstar\ (and SFR) is significant albeit with large scatter. Two coefficients
are calculated for our results: (1) Spearman's rank correlation coefficient
($r_s$) which measures the statistical dependence between two variables and (2)
Pearson product-moment correlation coefficient ($r$) which measures {\it
linear} correlation between two variables. The correlation between
H$\beta$--to--FUV ratio and \mstar\ (and SFR) is significant: $r_s$=0.39 (and
0.5), corresponding to a probability of $2.1\times10^{-7}$ (and
$4.3\times10^{-7}$) for the null hypothesis (no correlation) to be accepted.
The linear correlation test shows $r$=0.38 (and 0.47) between H$\beta$--to--FUV
ratio and \mstar\ (and SFR), indicating a modest {\it linear} relation. A
bootstrapping test shows that $r$ of both \mstar\ and SFR has a 5$\sigma$
significance level. We, therefore, fit a linear relation between
H$\beta$--to--FUV ratio and \mstar\ and SFR in Figure \ref{fig:result}.

The large scatter of the results in Figure \ref{fig:result} may be attributed
to one or more of the following causes: (1) measurement uncertainties of both
H$\beta$ and FUV, (2) galaxy-to-galaxy variation in attenuation/extinction
curves (in Section \ref{ratio}, we assume the same extinction curve for all our
galaxies), and/or (3) galaxies being observed at different times after the
onset of their temporary SF quenchings.

We also compare our results with those of W12 at $z=0$. W12 measured the flux
ratio between H$\alpha$ and FUV after correcting for dust extinction. We
convert their flux ratios to SFR ratios to obtain a direct comparison with our
results. The data of W12 show a ratio near unity from \mstar$\sim
10^{10.5}$\msun\ to \mstar$\sim 10^{8.5}$\msun. Their ratio starts to become
smaller than unity at even lower-mass ranges and eventually reaches 0.7 at
\mstar$\sim 10^{7}$\msun, while our ratio already reaches 0.7 at \mstar$\sim
10^{8.5}$\msun. This difference suggests the existence of redshift evolution of
the H$\beta$--to--FUV ratio for a given \mstar.

The correlation between H$\beta$--to--FUV ratio and SFR in our sample is also
different from that of \citet{leejanice09} and W12, after converting their flux
ratios to SFR ratios. At $z=0$, both \citet{leejanice09} and W12 show a
constant H$\alpha$--to--FUV ratio for galaxies with SFRs ranging from
0.1\msun/yr to 3\msun/yr. Their ratio decreases to 0.7 for galaxies with
SFR$\sim$0.01\msun/yr, which is about 50 times smaller than the SFR at which
the ratio of our sample reaches the same H$\beta$--to--FUV ratio. The redshift
evolution of the H$\beta$--to--FUV ratio shown as a function of \mstar\ is also
evident as a function of SFR.

Our sample covers a wide range of redshift, because we prefer to include as
many galaxies as possible to increase the robustness of the statistics. The cosmic
time interval of our sample's redshift range is about $\Delta t$=3.4 Gyr,
comparable to that from $z\sim0.4$ to $z\sim0$ ($\Delta t$=4.3 Gyr). Therefore,
there may also be redshift evolution of the H$\beta$--to--FUV ratio {\it
within} our sample. To test it, we divided our sample into two redshift bins to
calculate the H$\beta$--to--FUV ratios: $z=0.4-0.65$ ($\Delta t$=1.7 Gyr) and
$z=0.65-0.95$ ($\Delta t$= 1.5 Gyr). We only conduct this test for the \mstar\
range of ${\rm 10^9}$\msun--${\rm 10^{9.5}}$\msun. Above this mass range, the
signal of burstiness is almost zero. Below this mass range, our sample is
dominated by galaxies with $z<0.65$ because of the faintness of low-mass
galaxies.

The median H$\beta$--to--FUV ratios are log(H$\beta$/FUV)=-0.078 at
$z=0.4-0.65$ (38 galaxies with a median redshift of 0.51) and
log(H$\beta$/FUV)=-0.110 at $z=0.65-0.95$ (17 galaxies with a median redshift
of 0.79). The cosmic time interval between the medians of the two redshift bins
is 1.67 Gyr. The decreasing rate of log(H$\beta$/FUV) from the higher-redshift
bin to the lower one is, therefore, (0.110-0.078)/1.67 = 0.019 dex / Gyr. This
rate of redshift evolution {\it within} our sample is consistent with that
between our full \mstar=${\rm 10^9}$\msun--${\rm 10^{9.5}}$\msun\ sample (with
a median log(H$\beta$/FUV)=-0.096 and a median redshift of 0.53, corresponding
to a lookback time of 5.24 Gyr) and $z=0$ (e.g., W12).

We emphasize that the ``redshift evolution'' discussed above is for galaxies at
the same \mstar. We do not trace the mass evolution of the galaxies in our
sample, because we are not comparing the progenitors and descendants. Although
some individual nearby galaxies' mass assembly history can be determined (e.g.,
from \citet{harris09}, the LMC increased in log(\mstar) by ~0.4 dex between
$z\sim1$ and $z\sim0$), the knowledge of the average assembly history of
low-mass galaxies beyond the local universe is still lacking, which prevents us
from comparing the burstiness between progenitors and descendants. 

The comparison between galaxies with the same \mstar\ but different redshifts
has its own physical motivations, although it cannot trace the evolution of a
population. First, the SFR--\mstar\ relation increases with redshift. At the
same \mstar, higher-redshift galaxies have higher SFR and hence stronger
supernova feedback. Second, since the \mstar--halo mass relation evolves little
with redshifts \citep{behroozi13}, the same \mstar\ at different redshifts
corresponds to the same halo mass. But at a given halo mass, the cosmic
accretion rate increases with redshift \citep{dekel09gas}. Therefore, for the
same \mstar, higher-redshift galaxies have higher baryonic accretion rate. As
we discussed in Section \ref{intro}, strong feedback and a high gas accretion
rate are two conditions, which enable the periodic temporary quenching and new
accretion bursty cycles. Therefore, at the same \mstar, higher-redshift
galaxies are expected to be burstier, which is consistent with their observed
lower H$\beta$--to--FUV ratio.

In a recent paper, \citet{zeimann14} used the G141 grism of 3D-HST
\citep{brammer123dhst} to study the H$\beta$--to--FUV ratio in galaxies at
$1.90 < z < 2.35$. Their H$\beta$ SFR is a factor of $\sim$1.8 higher than that
expected from the systems' rest-frame UV flux density (i.e., log(H$\beta$/FUV)
$\sim0.25$). The apparent contradiction between their results and ours can be
attributed to two factors: metallicity and age. The sample of \citet{zeimann14}
is brighter than F140W$\sim$25.5 AB and has a median F140W magnitude of
$\sim$24.5 AB (roughly corresponding to \mstar\ $\sim10^{9.75} M_\odot$ at
their redshifts). We, therefore, only compare their results with our
intermediate and massive regimes. First, the sample of \citet{zeimann14} has a
metallicity distribution ranging from 12+log(O/H)$\sim$7.2 to $\sim$8.5 with a
median of $\sim$8.1. Their median is about 0.6 dex lower than the median
metallicity of $10^{9.75} M_\odot$ galaxies at $0.5<z<1.0$ \citep{ycguo16mzr}.
The formulae used to infer the SFR of local galaxies from H$\beta$ and FUV
measurements need to be modified for high-redshift low-metallicity galaxies, as
argued by \citet{zeimann14}. Second, the local formulae assume a system age of
100 Myr. The theoretical H$\beta$--to--FUV ratio changes little, if any, with
age when the age is older than 100 Myr, but the ratio increases quickly as age
decreases when the age is younger than 100 Myr. Therefore, applying the local
formulae for the galaxies in our sample (with a median age of 1 Gyr) is
reasonable, but the formula need to be modified for the high-redshift
strong-H$\beta$ galaxies of \citet{zeimann14}, which are likely younger than
100 Myr.

\section{Discussion}
\label{discussion}

There are four possibilities to explain the low H$\beta$--to--FUV ratio of
low-mass and/or low-SFR galaxies: (1) sample selection, (2) non-universal IMF,
(3) stochastic SF on star or star cluster scales, and (4) bursty SFH on
galactic scales. We will show that (1), (2), and (3) are not able to plausibly
explain the observed ratios alone, and, therefore, (4) is also needed. We also
add a detailed discussion on the effect of dust extinction correction in this
section.

\subsection{Sample Selection Effect}
\label{discussion:bias}

One of our main sample selection criteria is H$\beta$ S/N$>$3, which biases our
sample against H$\beta$-faint galaxies and, therefore, would artificially
increase the H$\beta$--to--FUV ratio. A second bias, however, is introduced by
our UV cut of F275W S/N$>$3 (Figure \ref{fig:sample}), which biases our sample
against UV-faint galaxies and would artificially decrease the H$\beta$--to--FUV
ratio. Which of the two biases dominates determines the actual systematic bias
in our observed H$\beta$--to--FUV ratio. 

Figure \ref{fig:bias} plots the H$\beta$--to--FUV ratio as a function of both
F275W magnitude and H$\beta$ flux. For galaxies with \mstar$<10^{9.5}$\msun,
their median H$\beta$--to--FUV ratio is almost independent of their F275W
magnitudes (blue in the left panel), but decreases with their H$\beta$ fluxes
(blue in the right). This result suggests that the H$\beta$ flux limit is the
main factor that biases our results: if we had a deeper spectroscopic sample to
extend the trends in the right panel of Figure \ref{fig:bias} to lower H$\beta$
flux, we would obtain an even lower median H$\beta$--to--FUV ratio for the
low-mass galaxies. On the other hand, including fainter UV galaxies from deeper
F275W images would not affect our results, because the median H$\beta$--to--FUV
ratio is almost independent of the F275W magnitudes. Therefore, our current
sample selection biases our H$\beta$--to--FUV ratio of low-mass galaxies
against the low-value side, implying that a complete sample of low-mass
galaxies might have even lower intrinsic H$\beta$--to--FUV ratios than what we
find.

\begin{figure*}[htbp] 
\includegraphics[scale=0.3, angle=0, clip]{./fig5a_proof.ps}
\includegraphics[scale=0.3, angle=0, clip]{./fig5b_proof.ps}

\caption[]{Comparisons between our observed H$\beta$--to--FUV ratio (as a
function of total SFR) and the predictions of two IGIMF models ({\it left}) and
the SLUG simulations ({\it right}). In both panels, black circles with gray
error bars are the best-measured ratios and their uncertainties from our
observations. Red squares with error bars show the median and 16th and 84th
percentiles of the black circles in each SFR bin. Blue symbols with error bars
in the left panel show the mean and the standard deviation of the mean of each
SFR bin.
Black solid and dotted lines show the best-fit linear relation and its 95\%
confidence level of the black circles. The predicted H$\beta$--to--FUV ratios
of the two IGIMF models are shown by the purple lines in the left panel. The
mean and 1$\sigma$ deviation of the H$\beta$--to--FUV ratios of the SLUG
simulations are shown by the brown squares with error bars in the right panel.

\label{fig:model}}
\end{figure*}

\subsection{Non-universal IMF}
\label{discussion:imf}

Typically, an IGIMF model assumes that (1) all stars form in clusters
\citep[e.g.,][]{pa07}; (2) the cluster mass function is a power law; (3) the
mass of the most massive star cluster (i.e., the truncation mass of the power
law) is a function of the total SFR of a galaxy; and (4) the mass of the most
massive star in a star cluster is a function of the mass of the cluster. These
assumptions result in a low H$\alpha$--to--FUV ratio by reducing the
probability of forming massive stars in low-SFR galaxies.  \citet{pa07,pa09}
presented the predicted H$\alpha$--to--FUV ratio as a function of
SFR$_{H\alpha}$ for a few IGIMF models. Here, we compare two of their models to
our observations: {\it Standard} and {\it Minimal1} \citep[see Table 1
of][]{pa07}. The Standard and Minimal1 models differ in the assumed cluster
mass functions. The Minimal1 scenario uses a cluster mass function that follows
$dN/dM \propto M^{-2}$ over most of its mass range in accord with most
observational determinations of the cluster mass function (e.g., see the review
by \citet{krumholz14}). In contrast, the Standard model assumes a steeper
$dN/dM \propto M^{-2.35}$ slope, i.e., it assumes that the cluster mass
function has the same slope as the stellar IMF\null. The latter choice
generates a much stronger IGIMF effect, because it places more of the SF in
smaller clusters with more strongly truncated IMFs. We convert their predicted
H$\alpha$--to--FUV ratios from a function of SFR$_{H\alpha}$ to a function of
SFR$_{FUV}$ (equivalent to SFR$_{tot}$ as their models are dust free) and
compare them with our H$\beta$--to--FUV ratio as a function of SFR$_{tot}$. 

Figure \ref{fig:model} shows that the Standard IGIMF model significantly
underpredicts the ratio by a factor of three for all galaxies, because of its
severely reduced probability of forming massive stars. Moreover, the slopes of
both Standard and Minimal1 models are flatter than our observed results. The
difference of the slopes would be even larger, if our sample selection effect,
which preferentially excludes galaxies with low H$\beta$--to--FUV (Section
\ref{discussion:bias}), is taken into account. The Minimal1 model lies within
the 1$\sigma$ level of the {\it scatter} of our sample. Its prediction is close
to our observed ratios for galaxies with SFR $\sim$1\msun/yr, but it
underpredicts the ratio for higher-SFR galaxies. To statistically test the
acceptance (or rejection) of the Minimal1 model, we calculate the
goodness-of-fit (reduced $\chi^2$) of the Minimal1 model to the mean and the
standard deviation of the mean of our results (blue points and error bars in
Figure \ref{fig:model}). The reduced $\chi^2$ is $\sim$22, much larger than
unity, and hence rules out the Minimal 1 model with more than 5$\sigma$
confidence.

We, therefore, conclude that the two types of IGIMF models are unable to
provide a plausible explanation of our results.

Similar results have been reported by W12. They found that the Minimal1 model
matches their results at $z=0$ well for galaxies with \mstar$>10^{8}$\msun\
and/or SFR$_{H\alpha} > 0.01$\msun/yr, while the Standard model systematically
underpredicts the H$\alpha$--to--FUV ratio.

\subsection{Stochastic Star Formation on Small Scales}
\label{discussion:slug}

The stochasticity considered here consists of stochastic sampling of both the
IMFs and SFHs (see Section \ref{intro}). As shown by \citet{leejanice09} and
\citet{fumagalli11}, stochastic sampling of IMFs alone cannot reproduce the
entire range of the observed H$\alpha$--to--FUV ratio, because it predicts a
systematically high value for very low-SFR local galaxies.

To fully describe the stochasticity, the simulation code SLUG
\citep{slug1,slug2,slug3} synthesizes stellar populations using a Monte Carlo
technique to properly treat stochastic sampling including the effects of
clustering, IMF, SFH, stellar evolution, and cluster disruption. Using SLUG,
\citet{fumagalli11} showed that stochasticity is able to explain the observed
low H$\alpha$--to--FUV ratio in local dwarf galaxies. Here, we compare the SLUG
simulations with our results to test if stochastic SF can explain our observed
H$\beta$--to--FUV ratio at $0.4<z<1.0$. Among all SLUG models, we use the one
that has the most significant stochastic effects by assuming all stars are
formed in clusters.  Also, the SLUG simulations are run to $z$=0.

Figure \ref{fig:model} shows that the SLUG's prediction of H$\beta$--to--FUV
ratio (brown squares with error bars) is systematically higher than the median
of our observed values. SLUG provides predictions only for galaxies with
log(SFR)$<$0.5, while our sample contains only galaxies with log(SFR)$>$-0.5.
The range of SFRs covered by both the predictions and observations is narrow.
Nevertheless, the comparison suggests that the stochastic SF alone is not able
to explain the low H$\beta$--to--FUV ratio of low-SFR galaxies in our sample.  

SLUG's predictions match the results of local dwarf galaxies of W12 very well.
The good agreement between SLUG and local observations (see also
\citet{fumagalli11} for comparison with observations before W12) and the
systematic offset of SLUG from our results suggest that, compared to the local
universe, an extra source is needed to fully explain the observed
H$\beta$--to--FUV ratio at $z\sim0.7$.

\begin{figure*}[htbp] 
\includegraphics[scale=0.3, angle=0, clip]{./fig6a_proof.ps}
\includegraphics[scale=0.3, angle=0, clip]{./fig6b_proof.ps}

\caption[]{H$\beta$--to--FUV ratio as a function of sSFR. {\it Left}: sSFR is
calculated by using dust-corrected H$\beta$-derived SFR, where the dust
extinction is measured through the ratio of ${\rm SFR_{tot}}/{\rm SFR_{FUV}}$
(see the upper panel of Figure \ref{fig:dust}).
{\it Right}: sSFR is calculated by using ${\rm SFR_{tot}}$. In both panels,
black circles with gray error bars are the best-measured ratios and their
uncertainties from our observations. Red squares with error bars show the
median and 16th and 84th percentiles of the black circles in each sSFR bin. 

\label{fig:ssfr}}
\end{figure*}

\begin{figure}[htbp] 
\includegraphics[scale=0.3, angle=0, clip]{./fig7.ps}

\caption[]{Burstiness of the SFHs of our galaxies. The burstiness, $(D \times
A)/P$, is measured from SFH models of W12 (see Section \ref{discussion:bursty}
for details). Red squares and error bars show the burstiness and 1$\sigma$
confidence level of our sample. The result of W12 at $z=0$ is shown as the blue
line. 

\label{fig:bursty}}
\end{figure}

\subsection{Burstiness of Low-mass Galaxies}
\label{discussion:bursty}

Because neither non-universal IMF nor stochastic SF is able to fully explain
our data, we believe that a bursty SFH is needed for low-mass galaxies at
$z\sim0.7$. This speculation is supported by the systematically lower
H$\beta$--to--FUV ratio at $z\sim0.7$ than at $z=0$. The main driver of the
bursty SFH is gas expulsion due to feedback followed by new gas accretion (or
recycling). This driver would be more efficient at high redshift when the
cosmic gas accretion rate is high and when galaxy dynamic timescales are short.
Therefore, if low-mass galaxies are bursty, the H$\beta$--to--FUV ratio at high
redshift is expected to be lower than that at low redshift, which is consistent
with our data.

One additional piece of evidence of the bursty SFH is the relation between the
H$\beta$--to--FUV ratio and sSFR in Figure
\ref{fig:ssfr}. The H$\beta$--to--FUV ratio clearly increases with sSFR {\it
when sSFR is measured through a nebular emission line}, while the ratio is
almost constant over a wide sSFR range {\it when sSFR is measured through UV+IR
or dust-corrected UV}. The trend with H$\beta$ sSFR is consistent with galaxies
undergoing starbursts: at the onset of a starburst, nebular emission-line
luminosity increases faster than FUV luminosity, resulting in a high H$\beta$
sSFR and a high H$\beta$--to--FUV ratio; while at the end of the burst (or at
the onset of temporary quenching), emission-line luminosity decreases faster
than UV luminosity,
resulting in low H$\beta$ sSFR and a low H$\beta$--to--FUV ratio. The duration
of a burst is likely much less than $\sim$100 Myr because when the UV+IR or
dust-corrected UV sSFR (an indicator averaged over 100 Myr) is used, the trend
disappears and the H$\beta$--to--FUV ratio remains a constant over a wide range
of sSFR. \citet{kurczynski16} also found no statistically significant increase
of the intrinsic scatter in the SFR--\mstar\ relation at low masses at
$0.5<z<3.0$ when the SFRs are measured from SED-fitting of broadband
photometry, which traces SFR on timescales of 100 Myr. Their results also
indicate that if the bursts exist, their timescale is likely much less than
$\sim$100 Myr.

The importance of burstiness for low-mass galaxies can be investigated through
SFH models. W12 constructed a series of models of bursty SFHs to explain the
observed H$\alpha$--to--FUV ratio at $z=0$. We use their models to explain our
results.

Their models assume an underlying constant SFH with normalized SFR=1 and
superposed with several starbursts over a period of 500 Myr. The bursts are
characterized by three key parameters: the burst amplitude ($A$, the increasing
factor of SFR over the underlying constant SFR during the burst duration),
burst duration ($D$), and burst period ($P$). The value of $(D \times A)/P$
indicates the relative importance of the bursty SF phase on SFR compared to the
constant SF phase. 

As discussed in W12, it is the distribution of H$\alpha$--to--FUV ratios that
contains information about the burstiness of the galaxies. We, therefore,
calculate the burstiness of our sample by matching the H$\alpha$ (or
H$\beta$)--to--FUV ratio distributions. We divide all galaxies in our sample
into five \mstar\ bins, starting from ${\rm 10^{8.5}}$\msun\ and each spanning
0.5 dex. For each bin (called our bin), we try to find a \mstar\ bin in the W12
sample, where the W12 galaxies have the most similar H$\alpha$--to--FUV
distribution with the H$\beta$--to--FUV distribution of the galaxies in our
bin. Specifically, for each of our bins, we run the K-S test to compare its
H$\beta$--to--FUV distribution with the H$\alpha$--to--FUV distributions of a
series of mass bins of W12. This series of mass bins consist of many bins with
width of 1 dex.  The smallest bin starts from ${\rm 10^6}$\msun, and each of
the next bin increases its lower-mass limit by 0.2 dex. These bins are designed
to be non-exclusive and their widths are wider than that of our bin to provide
robust statistics.  The W12 bin that has the largest K-S test probability has
the most similar H$\alpha$ (or H$\beta$)--to--FUV to our bin. For example, the
\mstar\ bin of ${\rm log(M_*/M_\odot)}=[9.0, 9.5)$ in our sample has the
largest K-S probability (0.94) with the \mstar\ bin of ${\rm
log(M_*/M_\odot)}=[7.5,8.5)$ in W12. We then use the burstiness of the bin of
$[7.5,8.5)$ in W12 (interpolate from Table 4 of W12) as the burstiness of our
$[9.0, 9.5)$ bin. We repeat this calculation for all our bins and the results
are shown in Figure \ref{fig:bursty}. The error bars are measured from a
bootstrapping test, which only samples three quarters of the galaxies in each
of our bins and repeats the random sampling 100 times.

Figure \ref{fig:bursty} shows that the median of $(D \times A)/P$ in our sample
increases toward lower \mstar. The three parameters in $(D \times A)/P$,
however, are coupled. A high $(D \times A)/P$ can be caused by either true
starbursts (i.e., $P \gg D$ and $(D \times A) \gg  P$) or ``gasping'' SFHs
(i.e., a short decrease in the SFR from an otherwise constant rate with $A \gg
1$ and $P \lesssim D$). The best W12 models used in our figures all have $P \gg
D$ with an average $P \sim$250 Myr and $D$ of a few tens of megayears,
indicating a true ``burst'' model.

The most massive galaxies have the median $(D \times A)/P \sim 0$, suggesting
their SFH is basically constant over timescales of 500 Myr. Intermediate-mass
galaxies have median $(D \times A)/P \sim 1$, suggesting an almost equal
contribution of SFRs from both bursty and constant SF phases. For low-mass
galaxies, the median $(D \times A)/P > 1$, suggesting that most of the SF
occurs during the bursty phase.  At \mstar$\sim {\rm 10^{8.5} M_\odot}$, the
bursty phase contributes a factor of three more SF than the constant phase
does.

We also compare our results with those of W12 at $z=0$. Our burstiness $(D
\times A)/P$ is larger than theirs at a given \mstar. In their sample, the
bursty phase becomes comparable to the constant phase only with galaxies at
\mstar$<10^{8.5}$\msun. But in our sample, it occurs at \mstar$\sim
10^{9.5}$\msun. The difference again clearly shows the redshift evolution of
the burstiness: at a given \mstar, galaxies at higher redshift are burstier,
i.e, have more of their SF occurs during the bursty phases.

Our results of low-mass galaxies being burstier than massive galaxies also show
excellent agreement with that from the Feedback in Realistic Environments
(FIRE) simulations \citep{hopkins14fire}.  \citet{sparre15} measured both
H$\alpha$ and FUV SFRs of the FIRE galaxies and found that the
H$\alpha$--to--FUV ratio in FIRE decreases toward small \mstar.  The FIRE ratio
is unity at ${\rm 10^{10} M_\odot}$ and decreases to $\sim$0.7 at ${\rm
10^{8.5} M_\odot}$, matching our observed results surprisingly well.  The FIRE
galaxies exhibit order-of-magnitude SFR variations over timescales of $\sim$10
Myr. Consequently, low-mass galaxies can go through both quenched (in terms of
the 10-Myr averaged SFR) and starburst phases a few times within a 200-Myr
period. The FIRE galaxies, however, are at redshifts lower than ours: $z=$0.0,
0.2, and 0.4. The FIRE H$\alpha$--to--FUV ratio is also systematically smaller
than that of W12, indicating that the FIRE galaxies are slightly burstier than
the observed local galaxies. 

Our results are also consistent with other indicators of bursty SFH. A model of
\citet{forbes14mzr} suggests that the scatter in both \mstar--SFR and
\mstar--gas-phase metallicity relations is mainly governed by the dispersion of
the baryonic accretion rate and/or the dispersion of the \mstar--\mhalo\
relation. Therefore, the increase of scatter toward the low-mass regime in one
relation would be accompanied by an increase in the other. Our observed
H$\beta$--to--FUV trend implies that the scatter in the \mstar--SFR relation
would increase with the decreasing of \mstar.  Therefore, the scatter in the
\mstar--gas-phase metallicity relation should also increase toward the low-mass
regime. Indeed, \citet{ycguo16mzr} found such an increase of the scatter in the
\mstar--gas-phase metallicity relation toward low-mass regimes, which
reinforces the bursty SFHs of low-mass galaxies at $z \sim 0.7$.

\begin{figure*}[htbp]
\includegraphics[scale=0.29, angle=0, clip]{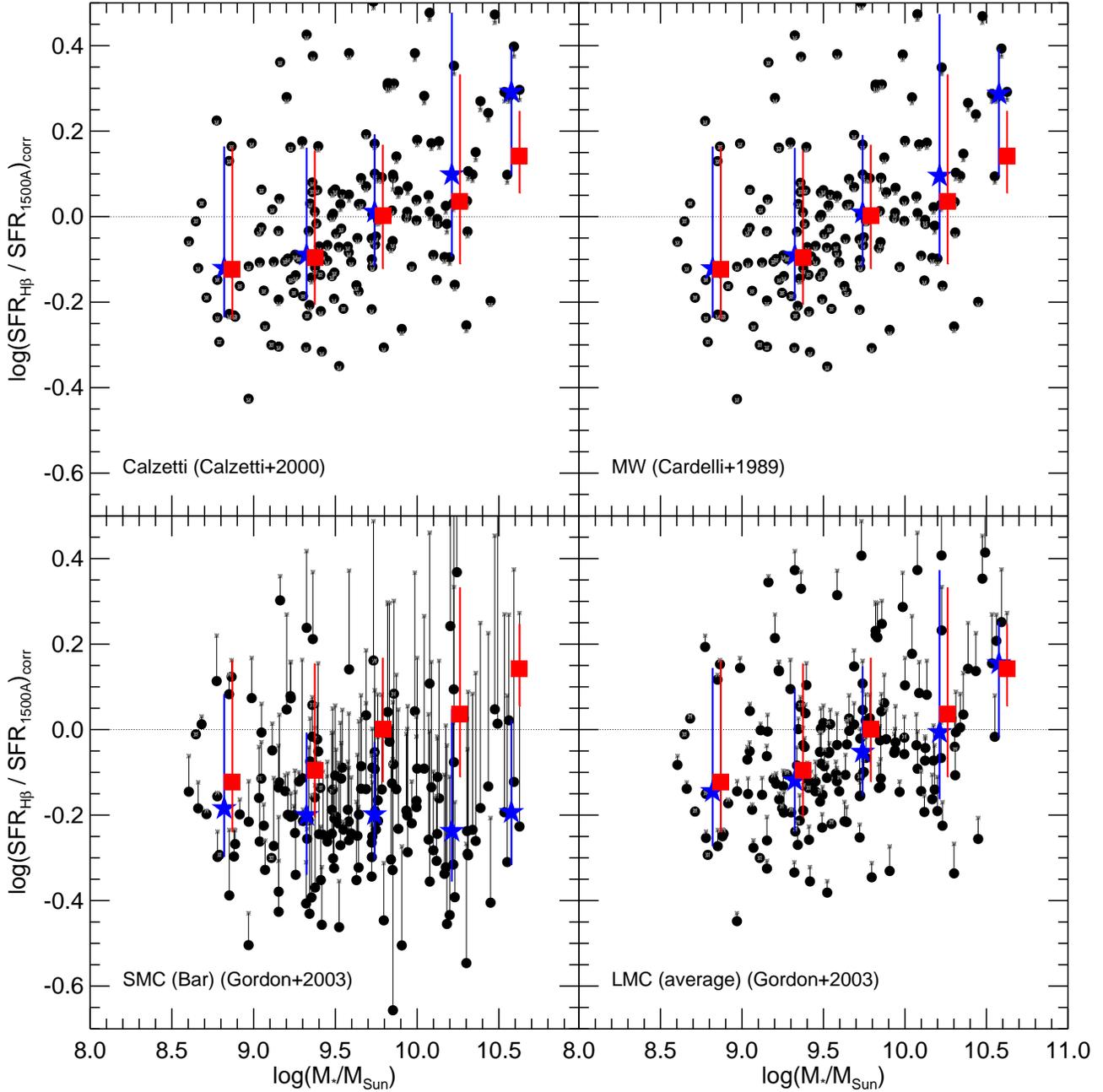}

\caption[]{Test of the H$\beta$--to--FUV ratios on the choice of the dust
attenuation curve. Each panel shows the {\it dust-corrected} H$\beta$--to--FUV
ratio as a function of \mstar, by using a certain dust attenuation curve as
indicated in the lower-left corner of each panel. Black circles show the {\it
dust-corrected} values, while gray stars connected to the black circles show
the {\it uncorrected} values. Blue stars with error bars show the median and
16th and 84th percentiles of the {\it dust-corrected} ratios (i.e., black
circles) of each \mstar\ bin. Red squares with error bars show the median and
16th and 84th percentiles of the original ratios of each \mstar\ bin by using
the shortcut of ignoring dust correction, which are the same as those in the
left panel of Figure \ref{fig:result}. 

\label{fig:avtest}}
\end{figure*}

\subsection{Effects of Dust Extinction Correction}
\label{discussion:dust}

Our ``shortcut'' of ignoring dust correction is enabled by the choice of the
Calzetti attenuation curve. 
The Calzetti curve was derived through local starburst galaxies. Because the
normalization of the SFR--\mstar\ relation increases with redshift, local
starburst galaxies would be the analogs of higher-redshift normal star-forming
galaxies, which makes the Calzetti curve possibly an appropriate and the most
commonly used one in the studies of distant (e.g., $z>0.5$) galaxies.
Choosing the Calzetti curve would make our paper consistent with other studies
of distant galaxies. For example, \mstar\ of the galaxies in our sample is
measured through SED-fitting with the Calzetti curve \citep{santini15}. The 
dependence of our results on the choice of the attenuation curve, however, 
needs to be discussed.

To this purpose, we measure the FUV--NUV color of each galaxy and then derive
its slope ($\beta$) of the rest-frame UV continuum. The UV slope provides a
measurement of the stellar extinction $E(B-V)_{stellar}$. We then derive the
gas extinction through $E(B-V)_{stellar} = 0.44 \times E(B-V)_{gas}$. With the
extinction of both gas and stellar components in hand, we then derive the
extinction correction factor $A(H\beta)-A(FUV)$, i.e., Equation (4), by using
four different attenuation curves: (1) Calzetti \citep{calzetti00}; (2) Milky
Way \citep{cardelli89}; (3) SMC Bar \citep{gordon03}; and (4) LMC average
\citep{gordon03}. The median and 16th and 84th percentiles of the {\it
dust-corrected} H$\beta$--to--FUV ratio are shown as blue stars and error bars
in Figure \ref{fig:avtest}. The red squares and error bars show our original
results of using the ``shortcut'' to ignore dust correction.

The new results with dust extinction corrected by using the Calzetti, Milky
Way, and LMC average attenuation curves are not significantly different from
our original results of ignoring dust. The tiny difference between the dust
corrected (black circles) and uncorrected (gray stars), especially for low-mass
galaxies, validates of our ``shortcut'' of ignoring the dust correction.
Therefore, our analyses and conclusions of the burstiness of low-mass galaxies
would not be significantly changed if these three curves are used. 

The only different result comes from the SMC curve, which changes the
H$\beta$--to--FUV ratio of massive galaxies more than that of low-mass galaxies
due to the higher dust extinction of massive galaxies. In this case, the
H$\beta$--to--FUV ratio is almost constant around -0.2 for all masses and seems
to match the prediction of the IGIMF Minimal1 model (the purple dotted line in
the left panel of Figure \ref{fig:model}). However, as shown by other studies,
the SFRs measured from FUV and nebular emission line are usually in very good
agreement for massive galaxies (see W12 for $z=0$ galaxies and
\citet{shivaei15,shivaei16} for $z\sim2$ galaxies).  Therefore, we suspect that
the SMC curve ``over-corrects'' the dust extinction for massive galaxies to
push their H$\beta$--to--FUV ratios to being significantly smaller than unity
(negative in logarithmic scale).

So far, we still have one assumption untested, namely $E(B-V)_{stellar} = 0.44
\times E(B-V)_{gas}$. The factor of 0.44 (called star--to--gas factor) is
argued to be a lower limit and the actual value may vary according to the
properties of the galaxies as well as the chosen attenuation curves. Here, we
test its relation with the attenuation curves. To this purpose, we assume that
any valid combination of the attenuation curve and the star--to--gas factor
would cause massive ($>10^{10} M_\odot$) galaxies to have the H$\beta$--to--FUV
ratio equal to unity. By minimizing the residual of the H$\beta$--to--FUV ratio
of massive galaxies to unity, we find that the best star--to--gas factor is
0.50 for the Calzetti curve, 0.51 for MW, 0.33 for SMC, and 0.48 for LMC.
Applying these values to each attenuation curve would not significantly change
our original results (i.e., low-mass galaxies have log(H$\beta$/FUV)
$\sim$-0.2). Our main conclusions of the burstiness of low-mass galaxies are
thus still valid.

In conclusion, we test different extinction curves and different reasonable
star--to--gas factors. Our main conclusions are hold for almost all of our
choices, expect for the SMC curve with the star--to--gas factor of 0.44.  It is
important to note that we always assume all galaxies have the same attenuation
curve and star--to--gas factor. There may be galaxy--to--galaxy variation. On
the other hand, however, none of our choices would increase the
H$\beta$--to--FUV ratio of low-mass galaxies to be significantly larger than
unity (positive in logarithmic scale). Therefore, even if the
galaxy--to--galaxy variation exists, our results of a smaller-than-unity median
H$\beta$--to--FUV ratio for low-mass galaxies are still robust.

Finally, another possible explanation of the observed H$\beta$--to--FUV ratio
has not been discussed in our paper: the loss of ionizing photons. Our
emission-line SFR calibration assumes that every emitted Lyman continuum photon
results in the ionization of a hydrogen atom. This assumption may be invalid
due to either of the two effects: (1) leakage of Lyman continuum photons into
the intergalactic medium and (2) absorption of Lyman continuum photons by dust
internal to the HII region. The loss of ionizing photons would result in a
lower-than-unity H$\beta$--to--FUV ratio. The two effects, however, may be
negligible for our low-mass galaxies. The Lyman continuum escape fraction is
less than 2\% for galaxies at $z\sim1$ \citep{siana07,siana10,rutkowski16}. And
the dust absorption would preferentially affect massive (and hence dustier)
galaxies rather than low-mass systems.

\section{Conclusions}
\label{conclusion}

We study the ratio of SFRs measured from H$\beta$ and FUV for galaxies at
$0.4<z<1$ in the CANDELS GOODS-N region by using the TKRS Keck/DEIMOS
spectroscopy and the newly available \hst/WFC3 F275W images from CANDELS and
HDUV. Our goal is to investigate the burstiness of the SFHs of low-mass
galaxies by using the H$\beta$--to--FUV ratio (SFR$_{H\beta}$/SFR$_{1500\AA}$).
Our sample contains 164 galaxies down to \mstar=$10^{8.5}$\msun. An advantage
of using H$\beta$ instead of H$\alpha$ is that the dust extinction effects on
H$\beta$ nebular line and on FUV stellar continuum (1500\AA) almost cancel each
other out, so that the dust extinction correction is negligible for most of our
galaxies when the {\it ratio} of H$\beta$ and FUV is measured.

We find that the H$\beta$--to--FUV SFR ratio increases with \mstar\ and SFR.
The median SFR$_{H\beta}$/SFR$_{1500\AA}$ ratio at \mstar$\sim10^{8.5}$\msun\
is lower than that at \mstar$\sim10^{10}$\msun\ by a factor of 0.7. In terms of
total SFR, the median H$\beta$--to--FUV ratio of galaxies with SFR$\sim
0.5$\msun/yr is about 0.7 times lower than that with SFR$\sim 10$\msun/yr. We
also find that at \mstar$<10^{9.5}$\msun, our median H$\beta$--to--FUV is lower
than that of local galaxies at the same \mstar, implying a redshift evolution.
Our sample selection biases our results against low H$\beta$--to--FUV ratios,
suggesting that the true H$\beta$--to--FUV ratio is even lower for a complete
sample of low-mass galaxies, which strengthens our results.

One model of non-universal IMF (IGIMF) scenario broadly matches our results for
galaxies with SFR$\sim 1$\msun/yr but cannot match the increase of the
H$\beta$--to--FUV ratio toward higher-SFR galaxies. Compared to our results,
SLUG simulations of stochastic SF on star cluster scales overpredicts the
H$\beta$--to--FUV ratio for low-SFR galaxies. 

Bursty SFHs provide a plausible explanation for the observed H$\beta$--to--FUV
ratios of low-mass galaxies at $z\sim0.7$. The burstiness increases as \mstar\
decreases. For galaxies with \mstar$<10^{9}$\msun, the SF burstiness is as
large as three, namely, within a period of $\sim$500 Myr, the amount of SF
occurring in starburst phases is three times larger than that in a smooth
continuous phase.  The burstiness of galaxies with \mstar$>10^{10}$\msun\ is
$<$1, namely, more stars formed in the smooth continuous phase than in bursty
phases. 

The bursty SF plays an important role in the assembly of low-mass galaxies.
Future work can improve our knowledge of it from three aspects: (1) deeper UV
imaging (e.g., UV Frontier Fields) and deeper spectroscopy (e.g., Halo7D) to
explore even lower \mstar\ regimes; (2) larger UV and emission-line surveys to
enlarge sample sizes; and (3) IR spectroscopy (e.g., MOSDEF) to investigate the
redshift evolution of the burstiness from higher redshifts.

\ \ \

We thank the anonymous referee for the valuable and constructive comments,
which improve this article. YG, SMF, DCK, GB, and HY acknowledge support from
NSF Grant AST-0808133.  Support for Program HST-GO-12060 and HST-AR-13891 were
provided by NASA through a grant from the Space Telescope Science Institute,
operated by the Association of Universities for Research in Astronomy,
Incorporated, under NASA contract NAS5-26555. MR acknowledges support from an
appointment to the NASA Postdoctoral Program at Goddard Space Flight Center.

{\it Facilities}: {\it HST} (WFC3), Keck (DEIMOS)

%
%



\bibliographystyle{apj}


\end{document}